\documentclass[aps,pre,twocolumn,superscriptaddress,groupedaddress]{revtex4}

\usepackage{latexsym}
\usepackage[dvips]{graphics}
\usepackage{amssymb,amstext,amsfonts,amsmath}

\usepackage[T1]{fontenc} 
\usepackage[utf8x]{inputenc}
\usepackage[usenames,dvipsnames]{color}

\usepackage{mhchem}

\usepackage{wasysym}
\usepackage{natbib} 
\date{\today}

\begin{document}


\title{Exactly solvable model of gene expression in proliferating bacterial cell population with stochastic protein bursts and protein partitioning}



\author{Jakub Jędrak}
\email[]{jjedrak@ichf.edu.pl}
\affiliation{Institute of Physical Chemistry, Polish Academy of Sciences, Kasprzaka 44/52, 01-224 Warsaw, Poland}

\author{Maciej Kwiatkowski}
\affiliation{unaffiliated; ORCID: 0000-0001-7675-4595}

\author{Anna Ochab-Marcinek$^{1}$}
%
%

\date{\today}

\begin{abstract}
\vspace{0.3cm}
%

Many of the existing stochastic models of gene expression contain the first-order decay reaction term that may describe active protein degradation or dilution. If the model variable is interpreted as the molecule number, and not concentration, the decay term may also approximate the loss of protein molecules due to cell division as a continuous degradation process. The seminal model of that kind leads to gamma  distributions of protein levels, whose  parameters are defined by the mean frequency of protein bursts and mean burst size. However, such models (whether interpreted in terms of molecule numbers or concentrations) do not correctly account for the noise due to protein partitioning between daughter cells. 

We present an exactly solvable stochastic model of gene expression in cells dividing at random times, where we assume description in terms of molecule numbers with a constant mean protein burst size. The model is based on a population balance equation supplemented with protein production in random bursts.  

If protein molecules are partitioned equally between daughter cells, we obtain at steady state the analytical expressions for probability distributions similar in shape to gamma distribution, yet with quite different values of mean burst size and mean burst frequency than would result from fitting of the classical continuous-decay model to these distributions. For random partitioning of protein molecules between daughter cells, we obtain the moment equations for the protein number distribution and thus the analytical formulas for the squared coefficient of variation. 



\end{abstract}
%
\maketitle

\section{Introduction}
Genetically identical cells in a proliferating microbial population may differ not only in their cell-cycle age, volume or growth rate, but also in the copy numbers of molecules, in particular, the protein molecules. Even if the dependence of gene expression on growth rate \cite{hintsche2013dilution, klumpp2014bacterial, keren2015noise, shahrezaei2015connecting} and cell-cycle stage \cite{walker2016generation} can be neglected, there still remain at least two factors causing fluctuations in the molecule numbers between individual cells: First, gene expression is an inherently stochastic process because of the small number of molecules involved in biochemical reactions. Second, molecules, including proteins, are partitioned between daughter cells at cell division in a more or less random manner \cite{huh2011non, huh2011random}. 


Yet, many {stochastic} models of gene expression neglect the protein partitioning  \cite{marathe2012deterministic}. The classical models with a first-order decay reaction term  \cite{paulsson2005models,cai2006stochastic, friedman2006linking, taniguchi2010quantifying} may be interpreted in two ways: (i) If the model describes the protein levels in terms of molecule numbers, then the degradation term describes not only a possible active protein degradation (rare in  bacteria \cite{Maurizi1992}) but it also mimicks the effect of cell division. However, this description neglects the fluctuations both due to the abrupt protein loss at cell division (which are present even at the half-by-half protein partitioning) and due to a possible random partitioning of the molecules between cells. (ii) If the same classical model describes the protein levels in terms of concentrations, then the continuous degradation term reflects, besides possible active degradation, the dilution due to cell growth. Under the assumption of a perfectly equal protein partitioning, the concentrations remain unchanged  at cell division. However, if protein partitioning is random, then it should result in concentration fluctuations (at binomial partitioning, they extinct at high gene expression levels), and these fluctuations are neglected in the conventional model. 
 
Based on the classical models of gene expression where proteins are assumed to be produced in random translational bursts \cite{paulsson2005models,cai2006stochastic, friedman2006linking, taniguchi2010quantifying} and on the population balance equations \cite{mantzaris2005cell, mantzaris2006stochastic, friedlander2008cellular}, we propose a minimal analytically tractable model of gene expression in a population of dividing cells. This model directly accounts for the protein  loss at cell division and random partitioning of the molecules between daughter cells.

The manuscript is organized as follows: Sec. \ref{sec:model} presents the theoretical model. In Sec. \ref{subsec:moments} we calculate the moments of the steady-state protein number distribution for any symmetric random partitioning of proteins between daughter cells. In Sec. \ref{subsec:noise_floor} we show that a \textit{noise floor}, i.e. the absolute lower bound for noise as a function of mean protein number, occuring for highly expressed proteins, arises in the model as a consequence of protein loss at cell division. This kind of a noise floor is a property of the description in terms of molecule numbers. In Sec. \ref{subsec:ours_vs_friedman} we compare the protein number distributions resulting from our model to the distributions from the seminal Friedman's model \cite{friedman2006linking}. In Sec. \ref{subsubsec:parameters_differ} we show that if both distributions have the same mean and variance then the underlying mean burst frequency  and mean burst size may differ. In Sec. \ref{subsubsec:analytical_distribution} we obtain the analytical form of the protein number distribution for the half-by-half protein partitioning and we show that its shape is very similar to the shape of a gamma distribution from the Friedman's model \cite{friedman2006linking}. In Sec. \ref{subsec:tails} we analyze the behavior of the distribution tails  for large protein numbers. In Sec. \ref{subsec:experimental_data} we compare the squared coefficient of variation of our model to the existing experimental data. Section \ref{sec:discussion} contains discussion and conclusions.


\section{Model}\label{sec:model}


We combine the analytical framework first introduced by Friedman et al. \cite{friedman2006linking}  with the population balance equation (PBE) \cite{mantzaris2005cell, mantzaris2006stochastic, friedlander2008cellular}, a standard tool for modelling microbial populations. 


In Ref. \cite{friedman2006linking}, gene expression was described as a compound Poisson process with random Poissonian events of protein production, where a random number of protein molecules was produced at once in each such event. The protein burst size $u$ was drawn from a distribution $\nu(u)$. Originally, $\nu(u)$ was assumed to be an exponential distribution \cite{friedman2006linking}, based on the experimental data for \textit{E. coli} \cite{yu2006probing,cai2006stochastic},  but here we will relax this assumption. Protein production was counterbalanced in the Friedman's model \cite{friedman2006linking} by a continuous, deterministic protein decay term. The model allowed self-regulation: The burst occurrence rate $k$ was modulated by a transfer function $h_p(x)$, dependent on the current protein level $x$. We note that the continuous variable $x$ was assumed to be protein concentration in Ref. \cite{friedman2006linking}. However, in this study, we interpret $x$ as the molecule number. This distinction is crucial because cell division exactly by half maintains a constant protein concentration, whereas the protein number drops by 1/2. If protein production bursts are constant in terms of molecule number, then their size in the units of concentration decreases in growing cells.

Therefore, we treat here the theoretical framework of Ref. \cite{friedman2006linking} as a continuous approximation to discrete fixed cell size gene expression models, cf. e.g. \cite{shahrezaei2008analytical}, and not in its original interpretation, as a model describing evolution of protein \textit{concentration} in growing and dividing cell.  
Within the model proposed in Ref. \cite{friedman2006linking}, the first order decay term describes both true protein degradation (if present) and protein dilution due to cell growth.

On the other hand, population balance equation (PBE) describes the time evolution of the cell number density in a proliferating microbial population \cite{mantzaris2005cell, mantzaris2006stochastic}. At cell division, protein molecules are partitioned according to $x\to \{qx, (1-q)x\}$, where {$x \in [0, \infty)$ is a continuous protein copy number, and not protein concentration, whereas} the division ratio $0 \leq q \leq 1 $ is a random number, drawn from the division ratio probability density function (PDF) $\eta(q)=\eta(1-q)$. 
Within the PBE approach as used in \cite{mantzaris2005cell, mantzaris2006stochastic, friedlander2008cellular}, protein production is described as a deterministic process. However, both protein production rate and cell division rate may depend on $x$ in a nontrvial way. 



By merging both the above theoretical frameworks, we get the time-evolution equation for $p(x,t)$, a PDF  of the protein number $x$  at time $t$ in the population of cells (Appendix \ref{Population_growth_rate_generation_time_distribution_age_structure}):

%
%
\begin{eqnarray}
\frac{\partial p(x,t)}{\partial t} &=& -\frac{\partial }{\partial x}\left[ g(x) p(x,t)\right]  - h_d(x) p(x,t) \nonumber \\ &+& 2 \int_{0}^{1} \frac{\eta(q)}{q} h_d\left(\frac{x}{q}\right) p\left(\frac{x}{q},t\right)dq  \nonumber \\ &-& p(x,t) \int_0^{\infty} h_d(\xi) p(\xi,t) d \xi  \nonumber \\
 &+& k \int_{0}^{x} w(x-x^{\prime}) h_p \left(x^{\prime} \right) p(x^{\prime},t) dx^{\prime}, 
\label{unregulated_t_dependent_ME_of_Friedman_with_protein_partition_PBE}
\end{eqnarray}  
%
{In the above equation}, $x-x^{\prime}$ is the burst size, $\nu(x-x^{\prime})$ is the burst size PDF, $w(x-x^{\prime}) \equiv \nu(x-x^{\prime}) - \delta(x-x^{\prime})$ and $\delta(x-x^{\prime})$ is Dirac delta distribution \cite{friedman2006linking}. The $x$-dependence of cell division and protein production rates are given by $h_d(x)$ and $h_p(x)$, respectively. Thus, $k h_p(x)$ is the burst frequency, whereas $g(x)$ describes either deterministic protein production or degradation. Eq. (\ref{unregulated_t_dependent_ME_of_Friedman_with_protein_partition_PBE}) is a generalization of both Eq. (1) of Ref. \cite{friedman2006linking} and Eq. (1) of Ref. \cite{friedlander2008cellular}. 

If $x$ cannot be treated as proportional to the cell mass or volume, it is unclear how the division rate should depend on the copy number of a given protein, and various functional forms of $h_d(x)$ were used in the literature \cite{mantzaris2005cell, mantzaris2006stochastic, friedlander2008cellular}. For that reason, and in order to obtain an  analytically tractable model, we focus here on the simplest case of the constant cell division rate, $2h_d(x) \equiv \Delta = const$, which implies $\int_0^{\infty} h_d(\xi) p(\xi) d \xi = \Delta/2$. 

\section{Results}
\subsection{Moments of the steady-state protein number distribution}\label{subsec:moments}

We consider the steady-state solution of Eq. (\ref{unregulated_t_dependent_ME_of_Friedman_with_protein_partition_PBE}), $\partial p(x, t)/\partial t = 0$. Assuming $g(x)=-\gamma x$, $\gamma \geq 0$ and unregulated protein production ($h_p(x) = 1$), we have 
\begin{eqnarray}
-\gamma \frac{d}{d x}\left[x p(x)\right] &=&  \Delta \int_{0}^{1} \eta(q)\left[\frac{1}{q} p\left(\frac{x}{q}\right) - p(x)\right] dq  \nonumber \\  &+& k \int_{0}^{x} w(x-x^{\prime})p(x^{\prime}) dx^{\prime}.
\label{unregulated_ME_Friedman_protein_partition_ss}
\end{eqnarray}
The Laplace transform ($\mathcal{L}\{\ldots\}$) of Eq. (\ref{unregulated_ME_Friedman_protein_partition_ss}) yields
\begin{equation}
\gamma s\frac{d G(s)}{d s} =  \Delta \int_{0}^{1} \eta(q) \left[ G(qs)- G(s)\right] dq  + k\hat{w}(s) G(s),
\label{unregulated_ME_Friedman_s_space_ss}
\end{equation}
where $G(s)=\mathcal{L}\{p(x)\}$, and $\hat{w}(s)=\mathcal{L}\{w(u)\}=\hat{\nu}(s)-1$. Moment equations follow from Eq. (\ref{unregulated_ME_Friedman_s_space_ss}), 
\begin{equation}
\left[(1-\mathcal{M}_r)\Delta + r\gamma \right] \mu_r =  k\sum_{l=1}^{r}\binom{r}{l}\mu_{r-l}m_l,
\label{equation_for_moments_of_p_ss}
\end{equation}
%
%
%
where $\mu_r = \int_{0}^{\infty} x^r p(x) dx$ is the $r$-th moment of the protein number PDF, $m_r = \int_{0}^{\infty} u^r \nu(u) du$ is the $r$-th moment of the burst size PDF, and $\mathcal{M}_r = \int_{0}^{1} q^r \eta(q) dq$ is the $r$-th moment of the division ratio PDF (note that $\mathcal{M}_1=1/2$).

For $\gamma \neq 0$, $\Delta = 0$ (the proteins are degraded, but there is no protein partitioning at cell division) Eq. (\ref{equation_for_moments_of_p_ss}) yields
\begin{equation}
\mu_1 = a m_1 \equiv \mu_{1,\gamma}, ~~~\kappa_2 = \mu_2 - \mu^2_1 = \frac{1}{2} a  m_2 \equiv \kappa_{2,\gamma},
\label{first_second_cumulant_p_a}
\end{equation}
where $a=k/\gamma$ \cite{jkedrak2016time}. On the other hand, for $\gamma = 0$ (the proteins are stable) and  $\Delta \neq 0$, from Eq. (\ref{equation_for_moments_of_p_ss}) we get 
\begin{equation}
\mu_{1} = 2 \omega m_1 \equiv \mu_{1,\Delta}, ~~~~  \kappa_{2} = \frac{4 \omega^2 \mathcal{M}_2 m^2_1 + \omega m_2}{(1-\mathcal{M}_2)} \equiv \kappa_{2,\Delta},   
\label{first_second_cumulant_p_omega}
\end{equation}
where $\omega=k/\Delta$. The variance of the protein number, $\kappa_{2,\Delta}=\text{var}[p(x)]$ (\ref{first_second_cumulant_p_omega}) is a monotonically increasing function of $(\mathcal{M}_2-\mathcal{M}_1^2)/\mathcal{M}_1^2 =  4\mathcal{M}_2-1$, i.e., of the noise due to protein partitioning.

For a given $\nu(u)$ and $\eta(q)$, if $a= 2\omega$, i.e., if $\mu_{1,\gamma}$ (\ref{first_second_cumulant_p_a}) is equal to $\mu_{1,\Delta}$ (\ref{first_second_cumulant_p_omega}), then from Eqs. (\ref{first_second_cumulant_p_a}) and (\ref{first_second_cumulant_p_omega}) it follows that $\kappa_{2,\Delta} > \kappa_{2,\gamma}$. Therefore, within the present description in terms of molecule numbers, the first-order decay such as in the classical model  by Friedman et al. \cite{friedman2006linking} is not equivalent to any protein partitioning mechanism.


\subsection{{Squared} coefficient of variation has a noise floor due to cell division}\label{subsec:noise_floor}

\begin{figure}
	\begin{center}					  				
		\scalebox{0.63}{\includegraphics{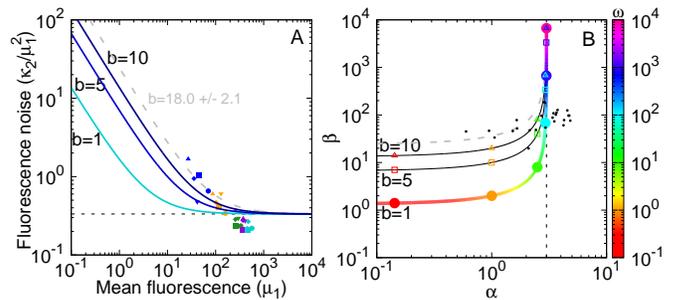}}
	\end{center}  
	\caption{(A) Plot of $c^2_v(\mu_1)$ (\ref{cv_squared_p_omega}) as a function of $\mu_1 = 2 \omega m_1$ for $m_2/m_1 = 2b$ and for selected mean burst sizes $b=m_1$.  Half-by-half protein partitioning is assumed, $\mathcal{M}_2 = 1/4$, which does not depend on $x$.  Dashed line: $\mathcal{E}_0=\mathcal{E}_{0, d}=1/3$, the lowest value of the noise floor allowed by the model. Points: Experimental data from Ref. \cite{Nordholt2017}, see Sec. \ref{subsec:experimental_data}). Symbols and colors in (A) are used as in Ref. \cite{Nordholt2017}. Symbols, carbon source: arabinose ($\CIRCLE$), maltose ($\blacksquare)$, glycerol ($\blacklozenge$), glucose  ($\blacktriangle$),  glucose$+$malate ($\blacktriangledown$). Colors, promoter activity induced by: 10 (blue), 20 (gold), 50 (green), 100 (violet), 1000 (turquoise) $\mu$M IPTG. Gray dashed curve: Fit of our model to the data. (B) Mapping of mean burst frequency $\omega$ and mean burst size $b$ of our model (where cell division is explicitly taken into  account) onto the corresponding values $\alpha$ and $\beta$ of these parameters appearing in Eq. (\ref{gamma_distribution_definition_alfabeta}) (where there is no protein loss at cell division but a continuous protein degradation is assumed). Curves: $\beta (\alpha)$ (\ref{eq_beta_as_function_of_alpha}) for variable $\omega$ and constant $b$. Colors denote the value of $\omega$. Circles, squares, and triangles denote $\omega = 10^i$, $i=-1,0,1,2,3,4$. Dashed line: upper bound for $\alpha$ within our model, $\alpha_{max}=\mathcal{E}^{-1}_{0, d}=3$. {Black dots: The same data as in (A) but the colors and symbols were omitted for the sake of clarity. Gray dashed curve: Fit from (A), mapped using Eq.  (\ref{eq_beta_as_function_of_alpha}).} } 
	\label{p_eta_u_sto}
\end{figure}

Protein degradation is important for eukaryotic organisms, but may be neglected in the case of bacteria;  hence we put $\gamma = 0$. From Eq. (\ref{first_second_cumulant_p_omega}) we obtain the squared coefficient of variation
\begin{equation}
c_v^2(\mu_1) \equiv \frac{\kappa_{2,\Delta}}{\mu_{1,\Delta}^2} = \frac{m_2}{2 m_1} \frac{1}{(1-\mathcal{M}_2)}  \frac{1}{\mu_1} + \frac{\mathcal{M}_2}{(1-\mathcal{M}_2)}.   
\label{cv_squared_p_omega}
\end{equation}
The protein {number} noise, {measured by the squared coefficient of variation} (\ref{cv_squared_p_omega}) is {thus} of the form  
\begin{equation}
c_v^2(\mu_1) = \frac{\mathcal{E}_1}{\mu_1} + \mathcal{E}_0,
\label{general_form_of_cv_2}
\end{equation}
%
%
with $\mu_1$-independent $\mathcal{E}_0$ and $\mathcal{E}_1$, which leads to the characteristic boomerang-like shape of the log-log plot (Fig. \ref{p_eta_u_sto}A). In Fig. \ref{p_eta_u_sto}, the half-by-half protein partitioning is assumed, 
	\begin{equation}
	\eta(q) = \eta_{d}(q)\equiv \delta\left(q-1/2\right),
	\label{half_by_half_eta}
	\end{equation}
	%
	%
which does not depend on $x$ and which gives $\mathcal{M}_2 = 1/4$. However, for small $x$, the half-by-half division as given by (\ref{half_by_half_eta}) is much less realistic than other continuous approximations to discrete binomial distribution, such as the 'all or none' partitioning ($\eta(q) = \eta_{b}(q)\equiv \frac{1}{2}\left[\delta (q) + \delta (1-q) \right]$) or even a uniform distribution of partition ratio ($\eta(q) =  \eta_{u}(q)\equiv\Theta(q)\Theta \left( 1-q \right)$), considered in Refs. \cite{brenner2007nonequilibrium,friedlander2008cellular}.

%
%
%
%
%


$\mathcal{E}_0=\mathcal{M}_2/(1-\mathcal{M}_2)$ provides the lower bound for extrinsic noise and does not depend on the details of protein production mechanism. However, the {contribution of protein partitioning to the value of} $\mathcal{E}_0$ predicted by the present model for a realistic level of protein partitioning noise is likely to be overestimated for the following reason: Eq. (\ref{unregulated_ME_Friedman_protein_partition_ss}) is a Master equation with time-independent parameters, hence it describes the cell division events as a Poisson process \cite{gardiner2009stochastic}. In consequence, the cell cycle length has an unrealistic, exponential distribution, contributing with too much noise (see Appendix \ref{generation_time_distribution}). Therefore, the age structure of the population is also unrealistic (Appendix \ref{cell_age_distribution}). This indicates a limitation of the simplest, single-variable version of the PBE approach as a candidate for a minimal model of gene expression as it assumes the "most random" distribution of cell cycle lengths (i.e., the one that naturally occurs from a Poisson process).

Thus, our model  provides  an "upper bound" for the noise contribution due to  cell cycle length fluctuations (meaning the "most random" fluctuations that can be generated by a model based on Master equation \cite{jafarpour2018bridging}): In the case when $\eta(q)$ is given by Eq. (\ref{half_by_half_eta}) there is no protein partitioning noise ($\text{var}[\delta \left(q- \frac{1}{2} \right)] = 0$) and $\mathcal{E}_0$ has the minimal value $\mathcal{E}_0 = 1/3 \equiv \mathcal{E}_{0, d}$. Therefore, $\mathcal{E}_{0, d}$ is the contribution to the noise floor coming solely from fluctuations of the cell cycle length, whereas the remaining part,  $\mathcal{E}_0 - \mathcal{E}_{0, d} = (4\mathcal{M}_2 - 1)/[3(1-\mathcal{M}_2)]$ is due to protein partitioning noise. \\

\subsection{Our model vs.  Friedman's model \cite{friedman2006linking}: Similar distributions but different  parameters}\label{subsec:ours_vs_friedman}

\subsubsection{Mean burst frequency  and mean burst size differ between the models despite the same  mean and variance of protein number distribution}\label{subsubsec:parameters_differ}
Let us consider a single-parameter burst size PDF, $\nu(u; b)=h\left(u/b \right)/b$, where $u \equiv x-x^{\prime}$ is a burst size, such that $m_1 = b $ and  $m_2 = C b^2$. $C=2$ for the exponential burst distribution: 

%
\begin{eqnarray}
\nu_{}(u) &=& b^{-1} e^{-u/b}, \nonumber \\ \hat{\nu}(s) \equiv  \mathcal{L}[\nu_{}(u)] & = & (sb+1)^{-1}.
\label{exponential_nu}
\end{eqnarray}
%

For $m_2/ m_1 = C b=const$, the increase in mean protein number $\mu_1=2\omega b$ is caused solely by the increase in $\omega$ (i. e., by the increase in burst frequency $k$ or decrease in cell division rate $\Delta$). In such a case, we move along the boomerang-shaped trajectories for a fixed value of  $b$, like those depicted in Fig. \ref{p_eta_u_sto}A. In general, $\mu_1$ may be varied both due to the change in $b$ and $\omega$. 

If $k=0$ and $g(x) = -\gamma x$ term in  Eq. (\ref{unregulated_ME_Friedman_protein_partition_ss}) is replaced with $g(x) = \sigma > 0$ (this is a special case of the model of Ref. \cite{friedlander2008cellular}), then instead of (\ref{cv_squared_p_omega}) we obtain 
\begin{equation}
c_v^2(\mu_1) = \mathcal{E}_0 = \frac{\mathcal{M}_2}{1-\mathcal{M}_2} = const.
\label{constant_coefficient_of_variation_for_deterministic_model}
\end{equation}
In other words, the squared coefficient of variation is then described by some general equation in the form of (\ref{general_form_of_cv_2}) (but not (\ref{cv_squared_p_omega})), where $\mathcal{E}_0 \neq 0$, $\mathcal{E}_1 = 0$ if protein production in our model is deterministic and cell division is stochastic. The opposite situation, $\mathcal{E}_1 \neq 0$, $\mathcal{E}_0 = 0$, is encountered for $\gamma \neq 0$, $\Delta = 0$, in particular for the model of unregulated, bursty gene expression with continuous protein degradation \cite{friedman2006linking}, which predicts that protein numbers are gamma-distributed:
\begin{equation}
p_{\gamma}(x; \alpha, \beta) \equiv \frac{x^{\alpha-1} \exp\left(-x/\beta \right)}{\beta^{\alpha}\Gamma(\alpha) }.
\label{gamma_distribution_definition_alfabeta}
\end{equation}
%
%
Then, $c_v^2(\mu_1) = 1/\alpha = \beta/\mu_1$, i. e., $\mathcal{E}_1=\beta$ and $\mathcal{E}_0=0$.
To make a comparison of $p_{\gamma}(x; \alpha, \beta)$ (\ref{gamma_distribution_definition_alfabeta}) {to} the protein number PDF of the present model, $p(x)$,  we assume now that these two distributions have equal means and equal second moments, $\mu_{r,\gamma} = \mu_{r,\Delta}$, $r=1,2$. We also assume that the burst size PDF $\nu(u)$ is given by (\ref{exponential_nu}), but the mean burst size may be different for the two models, $m_{1,\gamma} = \beta \neq b = m_{1,\Delta}$. These assumptions yield the following mapping: 
\begin{equation}
\alpha = \frac{\mu^2_1}{\kappa_2} = \frac{\omega\left(1-\mathcal{M}_2\right)}{\omega \mathcal{M}_2 + 1/2}, ~~~\beta  = \frac{\kappa_2}{\mu_1} = \frac{\left(2 \mathcal{M}_2 \omega + 1 \right)b}{1-\mathcal{M}_2}.  
\label{alpha_as_function_of_omega_0_ni_eta_EB}
\end{equation}
From (\ref{alpha_as_function_of_omega_0_ni_eta_EB}) we obtain
\begin{eqnarray}
\beta (\alpha) &=& \frac{b}{1 - (1 + \alpha )\mathcal{M}_2 }.  
\label{eq_beta_as_function_of_alpha}
\end{eqnarray}
For simplicity, we assume now that protein partitioning is deterministic, i.e., that each daughter cell obtains exactly half of the total number of protein molecules at cell division. In such a case, $\eta(q) = \eta_{d}(q)$ is given by Eq. (\ref{half_by_half_eta}), 
%
%
so that $\mathcal{M}_2 = 1/4$. In Fig. \ref{p_eta_u_sto}B we plot $\alpha$ vs.  $\beta$ (\ref{eq_beta_as_function_of_alpha}) to show that the mean burst frequency $\omega$ and mean burst size $b$ in our model take different values than the corresponding parameters $\alpha$ and  $\beta$ of the gamma distribution (\ref{gamma_distribution_definition_alfabeta}). In particular, $\omega$ is not limited to several bursts per cell cycle as was $\alpha$, so the mean burst size $b$ does not need to take as high values as would $\beta$ need to take in order to obtain a high level of gene expression.

\subsubsection{Analytical form of the protein number distribution for half-by-half protein partitioning. Apparent similarity to gamma distribution.}\label{subsubsec:analytical_distribution}

Below, we will show that equating the two first moments of our PDF $p(x)$ with those of the gamma distribution $p_{\gamma}(x)$ (\ref{gamma_distribution_definition_alfabeta}) yields similar overall shapes of the distributions. If so, any experimental gamma-shaped distributions with $c^2_v \geq \mathcal{E}_{0, d}=1/3$ may be as well fitted by the distributions given by our model (under the assumption that protein numbers, and not concentrations, were measured in experiment). With $\eta(q)=\delta(q-\frac{1}{2})$, we can rewrite Eq. (\ref{unregulated_ME_Friedman_s_space_ss}) as $G(s) = R(s) G\left(s/2\right)$, where $R(s)\equiv \{1 - \omega[\hat{\nu}(s) - 1]\}^{-1}$, $R(0)=1$. Solving by iteration, we get $G(s) = \prod_{i=0}^{\infty} R\left(s/2^{i} \right)$. For $\hat{\nu}(s)$ (\ref{exponential_nu}), $G(s)$ reads
\begin{eqnarray}
G(s) &=& \prod_{i=0}^{\infty} \frac{2^{-i}bs+1}{2^{-i}(1+\omega)bs+1} = \frac{(-b s;\frac{1}{2})_{\infty}}{(-b(1+\omega) s;\frac{1}{2})_{\infty}} \nonumber \\ 
&=& \sum_{r=0}^{\infty} \frac{((1+\omega)^{-1};\frac{1}{2})_{r}}{(\frac{1}{2};\frac{1}{2})_{r}} (-b(1+\omega) s)^r,
\label{unregulated_ME_Friedman_s_space_ss_etad_iloczyn_expburstsizepdf}
\end{eqnarray}
where $(a;q)_k$ is a $q$-Pochhammer symbol. In (\ref{unregulated_ME_Friedman_s_space_ss_etad_iloczyn_expburstsizepdf}) we used the $q$-binomial theorem \cite{andrews1986qseries},
\begin{equation}
\frac{(az;q)_{\infty}}{(z;q)_{\infty}} =
\sum_{n=0}^{\infty}\frac{(a;q)_n}{(q;q)_n} z^n. 
\label{q_Binomial_Theorem}
\end{equation}
The symbol $q$ appering in Eq. (\ref{q_Binomial_Theorem}) should not be confused with protein partitioning ratio $q$ appearing e.g. in Eqs. (\ref{unregulated_t_dependent_ME_of_Friedman_with_protein_partition_PBE}), (\ref{unregulated_ME_Friedman_protein_partition_ss}), (\ref{unregulated_ME_Friedman_s_space_ss}) or (\ref{half_by_half_eta}). The letter $q$ is traditionally used in the branch of mathematics called $q$-theory or $q$-analogs ($q$-binomial theorem or $q$-Pochhammer symbol are examples of such $q$-analogs).

\begin{figure}
\begin{center}					  				
\rotatebox{0}{\scalebox{0.6}{\includegraphics{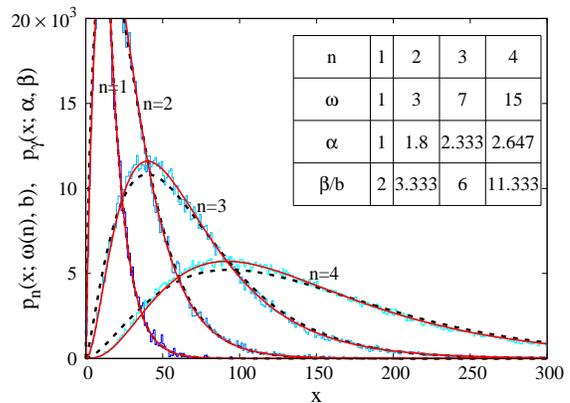}}} 
\end{center}  
\caption{Protein number PDFs of our model, $p_n(x)=p_n(x; \omega(n), b)$, vs. gamma distributions, $p_\gamma(x; \alpha,\beta)$.  Both distributions have similar shapes and become identical for $\omega = 1$ ($n=1$). Red solid line: $p_n(x)$ (\ref{p_d_finite_sum_specyfic_value_omega}) for $\omega = 2^n - 1$, $n=1, 2, 3, 4$ and $b=5$. Black dashed line: $p_\gamma(x; \alpha,\beta)$ (\ref{gamma_distribution_definition_alfabeta}) with $\alpha = 3 \omega/(\omega + 2)$ and $\beta/b  = 2 (\omega + 2) /3 $ given by (\ref{alpha_as_function_of_omega_0_ni_eta_EB}) for $\mathcal{M}_2 =1/4$. Blue lines: Simulations using the Gillespie algorithm (\cite{maarleveld2013stochpy}, see Appendix \ref{subsec:simulation}). Inset: Parameter comparison. } 
\label{p1_to_p4}
\end{figure}

From Eq. (\ref{unregulated_ME_Friedman_s_space_ss_etad_iloczyn_expburstsizepdf}) we obtain cumulants of $\mathcal{L}^{-1}\{G(s)\}$, 
%
\begin{equation}
\kappa_r = \frac{ 2^r b^r (r-1)! [(1+\omega)^r - 1]}{(2^r-1)}.
\label{cumulants_of_our_PDFs}
\end{equation}
We also have 
\begin{equation}
\mathcal{L}^{-1}\{G(s)\} = p(x) = \sum^{\infty}_{i=0} \frac{C_i(\omega)}{b} \exp\left(\frac{-2^{i}x}{b(\omega+1)} \right),
\label{p_d_as_infinite_sum}
\end{equation}
where 
\begin{equation}
C_{i}(\omega) = \frac{2^{i} \omega}{(1+\omega)^2}  \frac{(2(1+\omega)^{-1};2)_i}{(2;2)_i}\frac{(\frac{1}{2}(1+\omega)^{-1};\frac{1}{2})_{\infty}}{(\frac{1}{2};\frac{1}{2})_{\infty}}.
\label{D_i_p_d_as_infinite_sum_eta_d_iloczyn_stochastic_production}
\end{equation}
%
%
%
For  $\omega = \omega_n \equiv 2^n - 1$, $n=1, 2, 3, \ldots$, $ p(x)$ (\ref{p_d_as_infinite_sum}) can be written as a finite series 
\begin{equation}
p(x) \equiv p_{n}(x) = \frac{1}{b} \sum^{n}_{l=1} A_{n,l} \exp\left(-\frac{x}{2^l b } \right),
\label{p_d_finite_sum_specyfic_value_omega}
\end{equation}
%
%
where 
%
\begin{equation}
A_{n,l} = \frac{(-1)^{n-l} 2^{\frac{l(l-3)}{2}}}{\prod^{l-1}_{i=1} (2^i-1)\prod^{n-l}_{j=1} (2^j-1)}.  
\label{Definition_of_A_n_l}
\end{equation}
Equations (\ref{p_d_as_infinite_sum}) and (\ref{D_i_p_d_as_infinite_sum_eta_d_iloczyn_stochastic_production}) define a two-parameter family of PDFs, resembling the gamma PDF (Fig. \ref{p1_to_p4}). In particular, $p(x)$ (\ref{p_d_as_infinite_sum}) is right-skewed, unimodal for $\omega > 1$, and monotonically decreasing for $\omega \leq 1$. 
\subsection{Distribution tail for large protein numbers}\label{subsec:tails}
In both $p(x)$ (\ref{p_d_as_infinite_sum}) and $p_{\gamma}(x)$ (\ref{gamma_distribution_definition_alfabeta}), the mean burst size ($b$ and $\beta$, respectively) is the scaling parameter. The tail of $p(x)$ (\ref{p_d_as_infinite_sum}) is exponential: For large $x$ we have $p(x) \sim \exp[- x/(b(\omega+1))]$. However, in contrast to the gamma distribution, where the leading term is $ \exp[- x/b] $, the exponent in our model depends not only on the mean burst size, but also on burst frequency.

The same asymptotic behaviour as for $p(x)$ (\ref{p_d_as_infinite_sum}) is present if instead of $\eta(q)=\delta(q-\frac{1}{2})$ any other $\eta(q)$ is used (except for some pathological cases, e.g. for $\eta_{b}(q)=\frac{1}{2}\left[\delta (q) + \delta (1-q) \right]$) (Appendix \ref{Tails}). This is the special case of a yet more general result. Using Eq. (\ref{unregulated_t_dependent_ME_of_Friedman_with_protein_partition_PBE}), it can be shown in a similar manner as in Ref. \cite{friedlander2008cellular} that for $g(x)=0$ and $\nu(u)$ (\ref{exponential_nu}) the ratio $p(x_2)/p(x_1)$ at the tail of the PDF is well approximated by $\exp \left(-\mathcal{I}_{12} \right)$, where
\begin{equation}
\mathcal{I}_{12} = \int_{x_1}^{x_2}
\frac{k h^{\prime}_p(x) + h^{\prime}_d(x) + h_d(x)/b + R/b}{k h_p(x) + h_d(x) + R} dx,
\label{solution_general_ODE_tails}
\end{equation}
and $R=\int_0^{\infty} h_d(\xi) p(\xi) d \xi$ (Appendix \ref{Tails}). However, even if the tails are universal, i.e., independent on the division-ratio PDF $\eta(q) $, the corresponding probability distributions are not.  For example, for $\eta(q) =\Theta(q)\Theta \left( 1-q \right)$ used in Refs. \cite{friedlander2008cellular} and \cite{brenner2007nonequilibrium}, instead of (\ref{p_d_as_infinite_sum}) we obtain a statistical mixture of two broad ($c^2_v \geq 1/2$) gamma distributions (Appendix \ref{protein_PDF_for_eta_u}). Therefore in the present case it is no longer true that the ``steady-state population distribution (...) becomes insensitive to the division details'' \cite{brenner2007nonequilibrium}. For large $x$, $\eta(q) =\Theta(q)\Theta \left( 1-q \right)$ is much less realistic than $\eta(q) = \eta_{d}(q) = \delta(q-\frac{1}{2})$ (\ref{half_by_half_eta}) used here, and it leads to a higher noise floor ($\mathcal{E}_0 = 1/2$). 
\subsection{Comparison to experimental data}\label{subsec:experimental_data}
In this section, we compare the squared coefficient of variation as a function of the mean protein level in our model to the existing experimental data, under the assumption of equal protein partitioning between daughter cells (the lowest possible noise due to cell division, Fig.~\ref{p_eta_u_sto}A).

In Ref. \cite{Nordholt2017}, the authors measured total cell fluorescence emitted by the green fluorescent protein (GFP), encoded for by a  gene controlled by the hyper-spank promoter in \textit{B. subtilis}. The promoter activity was modulated by the concentration of  isopropyl-D-thiogalactoside (IPTG). An independent parameter varied in the experiment was the cell growth rate, depending on the carbon source in the medium. The data points shown in Fig.~\ref{p_eta_u_sto}A were obtained in Ref. \cite{Nordholt2017} by variation of these two parameters (we use the symbols and colors as in that Reference; we manually extracted the data points from Fig. 4a therein using the \textit{xyscan} software). 

Assuming that the total cell fluorescence scales linearly with the number of reporter proteins, the $x$ variable in our model may be reinterpreted as the fluorescence level and the mean protein burst size $m_1\equiv b$ is then expressed in fluorescence units. Thus, the mean protein number is proportional to the mean total cell fluorescence, and  the squared coefficient of variation does not depend on the units in which gene expression was measured (molecule number or fluorescence units). We note that the linear scaling between the GFP molecule number and total cell fluorescence is not necessarily an obvious assumption (see, e.g., Ref. \cite{newman2006single}). On the other hand, the authors of Ref. \cite{Nordholt2017} estimated that GFP maturation time did not affect strongly the fluorescence level, which may possibly exclude one source of non-linear scaling of the fluorescence detected vs. molecule number. The linear scaling was also assumed in Ref. \cite{Wolf2015}.

{The range of the mean total cell fluorescence measured in Ref. \cite{Nordholt2017} was too narrow to show a noise floor (Fig.~\ref{p_eta_u_sto}A). However, the data provide a hint that if the noise floor exists, it would be lower than that predicted by our model. This suggests that the distribution of cell cycle lengths in our model is indeed too wide, as we pointed out in Section \ref{subsec:noise_floor}.}

{For a rough comparison, we also plotted a fit of the squared coefficient of variation $c_v^2$ of our model (\ref{cv_squared_p_omega}) to the data. The slope of $c_v^2$ in our model seems to be in agreement with the experimental results. The fit yields an estimation of the mean burst size $m_1\equiv b = 18.0 \pm 2.1$ in the units of fluorescence, as used in Ref. \cite{Nordholt2017}; however, this value should be treated with caution because some other model with a lower noise floor due to a narrower distribution of cell-cycle lengths may possibly yield a different value of $b$.} 

{For completeness, we also plotted the data from Ref. \cite{Nordholt2017}  in Fig.~\ref{p_eta_u_sto}B (we omitted the colors and symbols used in Fig.~\ref{p_eta_u_sto}A). We made the assumption that the total cell fluorescence in the experiment was gamma-distributed, as shown in the SI of Ref. \cite{Nordholt2017}. We mapped the mean $\langle f \rangle$ and squared coefficient of variation $c_{v,f}^2=var(f)/\langle f \rangle^2$ of the total cell fluorescence onto the corresponding values of parameters of gamma distributions: $\alpha = (c_{v,f}^2)^{-1}$, $\beta = \langle f \rangle \cdot c_{v,f}^2$. }

{A noise floor was observed in other experiments reported in literature \cite{volfson2006origins,keren2015noise,taniguchi2010quantifying,Silander2012,Wolf2015,newman2006single, bar2006noise} but most of these data were not suitable for comparison to our model (see an overview in Appendix \ref{app:other_data}) because they were normalized to cell volume or gated to make them independent on cell-cycle stage. Thus, molecule number fluctuations due to loss of proteins at cell division were absent in these data. }

{There are two studies on \textit{S. cerevisiae} that reported the gene expression noise floor in ungated measurements. Its levels lower than in our model (where $\mathcal{E}_0=1/3$): In Ref. \cite{volfson2006origins}, Fig. 2 and S2 therein, the noise floor measured as the coefficient of variation $c_v$ took the values between 0.3 and 0.4, corresponding to the squared coefficient of variation $c_v^2$ between 0.09 and 0.16. In Ref. \cite{keren2015noise}, Fig. 2A and S9 therein, $c_v^2$ was between 0.1 and 0.2. One reason for that difference may be the too wide distribution of cell cycle lengths in our model. But it should also be noted that our model may be far too simplistic for eukaryotic cells, as it does not account for the discrete  promoter activity states due to chromatin remodeling, the nuclear transport, etc.}
%
%

\section{Discussion and conclusions}\label{sec:discussion}

We have proposed a model of gene expression in a proliferating cell population, which is a generalization of the models of Refs. \cite{friedman2006linking} and Ref. \cite{friedlander2008cellular}. In Ref. \cite{friedman2006linking}, the protein production was stochastic, whereas the decrease of protein concentration was due to a deterministic protein degradation process. Here, we treat the model of Ref. \cite{friedman2006linking} as a continuous approximation (still in the units of protein number) to discrete gene expression models, such as in Ref. \cite{shahrezaei2008analytical}, that describe non-growing and non-dividing cells. 

In Ref. \cite{friedlander2008cellular}, the situation was just opposite: Protein partitioning was stochastic and  protein production was deterministic. After combining the stochastic protein production (of Ref. \cite{friedman2006linking}) with the stochastic protein partitioning (used in Ref. \cite{friedlander2008cellular}), we obtain the  boomerang-like shape of the log-log plot of mean protein copy number vs. protein copy number noise.

Because the protein \textit{copy number} and not protein \textit{concentration} is the variable in our model, the protein partitioning at cell division along with an age structure of the cell population (cf. Eq. (\ref{age_distribution_second_time_SI})) lead to the existence of the noise floor -- an absolute lower bound for noise, present for highly expressed proteins. 


 Our results suggest that the values of mean burst size and burst frequency that may be obtained by fitting theoretical distributions to experimental data are strongly model-dependent.  Na\"{i}ve fitting of the gamma distribution \cite{friedman2006linking}  or its discrete counterpart, negative binomial distribution \cite{shahrezaei2008analytical}, to the data measured in terms of molecule numbers, would neglect protein loss at cell division because the underlying models neglected cell division, and thus such a fitting might overestimate the mean burst sizes and underestimate mean burst frequencies for higher gene expression levels. 
 

Our model, directly accounting for the protein molecule numbers, may be important for interpretation of experimental results. In some studies, data correction for cell size  was carried out: In \cite{taniguchi2010quantifying}, cell volume was measured by image recognition and the protein fluorescence was normalized by the volume. In other studies, the flow cytometry data were gated \cite{Silander2012, Wolf2015, newman2006single, bar2006noise}. Gating filters out the data only for those cells that scatter light to a similar extent, so the observed cells are supposed to be of similar sizes. However, the gating procedure may be imperfect because setting the size range too narrow leaves too little data, whereas setting the gate too wide results in some variation in cell sizes. One can avoid these problems by measuring the total cell fluorescence \cite{Nordholt2017, volfson2006origins, keren2015noise} and then separating the noise contributions from varying cell sizes (due to cell-cycle progression and other sources of variability) using independent measurements \cite{Nordholt2017} or estimating these contributions theoretically. Our model brings us closer to theoretical determination of the magnitude of one of these contributions, namely, the protein loss due to cell division and partitioning between daughter cells. 
	
We have compared our model to the existing literature data \cite{Nordholt2017}, however, the possibilities of such a comparison are limited so far. The results for \textit{B. subtilis}, reported in Ref. \cite{Nordholt2017}, were measured in too narrow a range of gene expression levels, whereas the results for \textit{S. cerevisiae} of Refs. \cite{volfson2006origins,keren2015noise} may not be comparable to our minimal model because of the complexity of gene expression mechanisms in eukaryotes.
However, a cautious comparison of our model to the above data suggests that the main limitation of the model is its prediction of too large a contribution of protein loss at cell division to the noise floor. This is due to the exponential distribution of cell cycle lengths, since the cell divisions are modelled as Poissonian events (Appendix \ref{generation_time_distribution}). 
In a more realistic model, the cell cycle length distribution could be modeled, instead of the exponential distribution, as gamma distribution \cite{powell1956growth}  or some other distribution peaked around the mean cell cycle length, with the limiting case of Dirac delta distribution describing a deterministic cell cycle. However, this is beyond the scope of the present paper. In preparation is our new paper \cite{jedrakochabinpreparation} that explores the noise levels in a more realistic model.


Another limitation of our theoretical approach may be the  description of the messenger RNA (mRNA) as very short-lived molecules, only implicitly present in the model \cite{friedman2006linking}, and thus the neglect of their partitioning at cell division. Also, the model does not describe the factors that may significantly shape  the  gene expression noise levels in eukaryotes, e. g., the discrete on-off promoter switching or nuclear transport. 

We may expect that negative gene autoregulation (not considered here) would suppress protein noise \cite{friedman2006linking, jia2017stochastic}, and thus decrease the noise floor. Yet, positive autoregulation is expected to have the opposite effect \cite{friedman2006linking, jia2017stochastic}. In a similar manner, nontrivial dependence of cell division rate on protein number ($h_d(x)\neq const$) would probably result in a lower noise floor. Still, the analytical solution to our model in presence of gene autoregulation or cell division regulation does not seem feasible. For that reason, we have studied here the non-regulated gene expression and cell division only.

\section*{Acknowledgments}
AOM was supported by the National Science Centre SONATA BIS 6 grant no. 2016/22/E/ST2/00558.


\section*{Author contributions}

JJ designed the study, performed the analytical calculations and wrote the manuscript. MK wrote the simulation code for cell division. AOM participated in the study design by the idea of comparison of the model with experimental data and by making the comparison; designed the simulation algorithm for cell division, supervised writing the simulation code and introduced minor modifications in the code; performed the simulations; performed the distribution fitting by maxima matching and observed heuristically the behavior of the probablity distributions described in Appendix \ref{subsec:property}; participated in writing the manuscript.

\appendix 

\section{Derivation of Eq. (\ref{unregulated_t_dependent_ME_of_Friedman_with_protein_partition_PBE}) \label{Population_growth_rate_generation_time_distribution_age_structure}}
%
%

Equation (\ref{unregulated_t_dependent_ME_of_Friedman_with_protein_partition_PBE}) follows from an apparently very similar, but more fundamental equation describing the number density of cells, $F(x,t)$, in a proliferating population,
\begin{eqnarray}
\frac{\partial F(x,t)}{\partial t} &=& -\frac{\partial }{\partial x}\left[ g(x) F(x,t)\right] - h_d(x) F(x,t) \nonumber \\ &+& 2 \int_{0}^{1} \frac{\eta(q)}{q} h_d\left(\frac{x}{q}\right) F\left(\frac{x}{q},t\right)dq  \nonumber \\
 &+& k \int_{0}^{x} w(x-x^{\prime}) h_p \left(x^{\prime} \right) F(x^{\prime},t) dx^{\prime}.
\label{regulated_t_dependent_PBE_with_protein_bursts_SI_1}
\end{eqnarray}  
Namely, $F(x,t)dx$ is the number of those cells in a population, which at time $t$ contain exactly $x$ protein molecules. In order to derive Eq. (\ref{unregulated_t_dependent_ME_of_Friedman_with_protein_partition_PBE}) from Eq. (\ref{regulated_t_dependent_PBE_with_protein_bursts_SI_1}), we define
%
%
\begin{equation}
p(x,t) = \frac{F(x,t)}{\mathcal{N}(t)},
\label{chagne_of_variables_F_to_p}
\end{equation}
where 
\begin{equation}
\mathcal{N}(t) = \int_{0}^{\infty} F(x,t) dx
\label{definition_of_M}
\end{equation}
is the total number of cells in the population \cite{mantzaris2006stochastic}. 
Integrating both sides of Eq. (\ref{regulated_t_dependent_PBE_with_protein_bursts_SI_1}) from $x=0$ to $x=\infty$, assuming that $F(\infty, t) = g(0)F(0,t) = 0$ and making use of (\ref{chagne_of_variables_F_to_p}) and (\ref{definition_of_M}) we obtain
\begin{equation}
\frac{d \mathcal{N}(t)}{d t} = \mathcal{N}(t) \int_{0}^{\infty} h_d(x) p(x,t) dx.
\label{population_growth_from_Mantzaris}
\end{equation}

Now, Eq. (\ref{unregulated_t_dependent_ME_of_Friedman_with_protein_partition_PBE}) follows from (\ref{regulated_t_dependent_PBE_with_protein_bursts_SI_1})-(\ref{population_growth_from_Mantzaris}) and from obvious identity
\begin{equation}
\frac{\partial F(x,t)}{\partial t} = \mathcal{N}(t) \frac{\partial p(x,t)}{\partial t} + p(x,t) \frac{d \mathcal{N}(t)}{d t}.
\end{equation}
The above derivation is essentially that of Ref. \cite{mantzaris2006stochastic}, the only difference is the presence of the terms responsible for bursty protein production in the present case.   

In spite of their apparent similarity, there are important differences between Eq. (\ref{unregulated_t_dependent_ME_of_Friedman_with_protein_partition_PBE}) and Eq. (\ref{regulated_t_dependent_PBE_with_protein_bursts_SI_1}). First, the terms related to protein partitioning in Eq. (\ref{regulated_t_dependent_PBE_with_protein_bursts_SI_1}) conserve the total number of protein molecules (molecules are neither created, nor destroyed during cell division), whereas from Eq. (\ref{unregulated_t_dependent_ME_of_Friedman_with_protein_partition_PBE}) it follows that the  cell division always decreases the mean protein number in the population. This is due to the $(-1)p(x,t) \int_0^{\infty} h_d(\xi) p(\xi,t) d \xi $ term in Eq. (\ref{unregulated_t_dependent_ME_of_Friedman_with_protein_partition_PBE}), not present in Eq. (\ref{regulated_t_dependent_PBE_with_protein_bursts_SI_1}). 

Second, $\mathcal{N}(t)$ grows indefinitely according to Eq. (\ref{population_growth_from_Mantzaris}), therefore in contrast to Eq. (\ref{unregulated_t_dependent_ME_of_Friedman_with_protein_partition_PBE}), Eq. (\ref{regulated_t_dependent_PBE_with_protein_bursts_SI_1}) does not have nontrivial stationary solutions. 

Our task now is to determine, for the present model: (i) the population growth rate, $\nu_m $, (ii) probability distribution $f(\tau)$ of generation time (cell cycle length), $\tau$, for each newborn cell, and (iii) the cell age ($\tilde{a}$) distribution $\phi(\tilde{a})$ in the state of a balanced, exponential growth \footnote{In this section we use notation of Ref. \cite{powell1956growth}, but the cell age is denoted $\tilde{a}$ and not $a$ to avoid confusion with quantity $a = k/\gamma$ defined in the main text.}. This is done in next three sections.

\section{Population growth rate $\nu_m$ \label{growth_rate}}

The steady state solutions of Eq. (\ref{unregulated_t_dependent_ME_of_Friedman_with_protein_partition_PBE}), $p(x)$, correspond to the so called state of balanced growth, when the shape of $F(x,t)= \mathcal{N}(t)p(x)$ does not change but is only rescaled by $\mathcal{N}(t)$ \cite{mantzaris2006stochastic}. In such a case, from Eq. (\ref{population_growth_from_Mantzaris}) we obtain  
\begin{equation}
\mathcal{N}(t) = \mathcal{N}_0 e^{\nu_m t},
\label{population_growth_solution_SI}
\end{equation}
where $\nu_m$, given by  
\begin{equation}
\nu_m =  \int_{0}^{\infty} h_d(x) p(x) dx
\label{population_growth_rate_Mantzaris}
\end{equation}
is the population growth rate. $\mathcal{N}(t)$ as given by (\ref{population_growth_solution_SI}) is characteristic for (in fact, defines) the phase of exponential growth \cite{powell1956growth}\footnote{We assume that the number of cells in the population, $\mathcal{N}(t)$, is sufficiently large its fluctuations may be neglected, and that $\mathcal{N}(t)$ may be regarded as evolving according to deterministic equation (\ref{population_growth_solution_SI}).}.  

If $g(x)=-\gamma x$ and if neither the protein production rate nor the cell division rate depend on the number of protein molecules ($h_p(x)=1$, $h_d(x) \equiv \Delta/2$), Eq. (\ref{unregulated_t_dependent_ME_of_Friedman_with_protein_partition_PBE}) reads  
\begin{eqnarray}
\frac{\partial p(x,t)}{\partial t} &=& \gamma \frac{\partial}{\partial x}\left[x p(x,t)\right] + k \int_{0}^{x} w(x-x^{\prime})p(x^{\prime},t) dx^{\prime} \nonumber \\  &+&  \Delta \int_{0}^{1} \eta(q)\left[\frac{1}{q} p\left(\frac{x}{q},t\right) - p(x,t)\right] dq.
\label{unregulated_ME_Friedman_protein_partition_time_dependent_SI}
\end{eqnarray}
Central to this paper is the steady-state limit of Eq. (\ref{unregulated_ME_Friedman_protein_partition_time_dependent_SI}), 
\begin{eqnarray}
-\gamma \frac{d}{d x}\left[x p(x)\right] &=&  \Delta \int_{0}^{1} \eta(q)\left[\frac{1}{q} p\left(\frac{x}{q}\right) - p(x)\right] dq  \nonumber \\  &+& k \int_{0}^{x} w(x-x^{\prime})p(x^{\prime}) dx^{\prime},
\label{unregulated_ME_Friedman_protein_partition_ssSI}
\end{eqnarray}
i.e., Eq. (2). Although time-independent, Eq. (\ref{unregulated_ME_Friedman_protein_partition_ssSI}) describes the stationary protein distribution $p(x)$ in a \textit{growing} population of dividing cells (i.e., the state of balanced growth), as discussed above. 
%
%
In particular, for $h_d(x) = \Delta/2 \equiv \Delta_0$, from Eq. (\ref{population_growth_rate_Mantzaris}) it follows that 
\begin{equation}
\nu_m=\Delta_0, 
\label{growth_rate_equals_division_rate}
\end{equation}
%
%
i.e., the population growth rate $\nu_m$ appearing in Eq. (\ref{population_growth_solution_SI}) is equal to individual cell division rate $\Delta_0$, as should be expected.

It is instructive to show this result in an alternative way. Consider time evolution of the moments of $p(x,t)$, resulting from Eq. (\ref{unregulated_ME_Friedman_protein_partition_time_dependent_SI}), 
\begin{equation}
\dot{\mu}_r = - \left[(1-\mathcal{M}_r)\Delta + r\gamma \right] \mu_r +   k\sum_{l=1}^{r}\binom{r}{l}\mu_{r-l}m_l.
\label{equation_moments_time_dependent_SI}
\end{equation}
In the absence of protein production ($k=0$) and degradation ($\gamma=0$), from Eq. (\ref{equation_moments_time_dependent_SI}) it follows that the time dependence of the first moment of $p(x,t)$, i.e., the mean protein number is given by \footnote{In such a case the only possible steady state is a trivial one, i.e., $\lim_{t \to \infty} p(x, t) = p(x)=\delta(x)$.} 
\begin{equation}
\mu_1(t) = \mu_1(0)e^{-\Delta_0 t}.
\label{evolution_mean_number_no_production_SI}
\end{equation}
On the other hand, from the definition of the population average, the mean protein number $\mu_1(t)$ is equal to $X(t)/\mathcal{N}(t)$, where $X(t) \equiv \sum_{i=1}^{\mathcal{N}(t)}x_i(t)$ denotes the total number of protein molecules in a population, and $x_i(t) $ is a number of protein molecules in $i$-th cell at time $t$. If there is no protein production or degradation, $X(t)$ is constant and equal to its initial value, $X(t) = X(0) = \mathcal{N}_0 \mu_1(0)$. In such a case, time evolution of the mean protein number $\mu_1(t)$ is caused solely by the increase in cell number,
%
\begin{equation} 
\frac{X(t)}{\mathcal{N}(t)} = \frac{\mathcal{N}_0 \mu_1(0) }{\mathcal{N}_0 e^{\nu_m t}} = \mu_1(0)e^{-\nu_m t}.
\label{mean_protein_number_from_definitions_SI}
\end{equation}
%
%
Comparing (\ref{evolution_mean_number_no_production_SI}) and (\ref{mean_protein_number_from_definitions_SI}) we see that $\nu_m = \Delta_0$, i.e., the population growth rate $\nu_m$ is equal to the individual cell division rate $\Delta_0$, as it should.
%
%
%
%

So far, we have considered the whole proliferating cell population, for which $\Delta = 2 \Delta_0 = 2\nu_m$ (scenario A). However, Eq. (\ref{unregulated_ME_Friedman_protein_partition_time_dependent_SI}), but not Eq. (\ref{unregulated_t_dependent_ME_of_Friedman_with_protein_partition_PBE}), may be also used to describe the time evolution of the protein number distribution $p(x,t)$ in a single cell lineage. In such a case, we discard one of the daughter cells at each cell division, and therefore $\Delta = \Delta_0$ (scenario B). 



\section{Generation time distribution $f(\tau)$ \label{generation_time_distribution}}



In order to find the generation time distribution $f(\tau)$, consider Eq. (\ref{unregulated_ME_Friedman_protein_partition_time_dependent_SI}). This is a special case of the differential Chapman-Kolmogorov equation \cite{gardiner2009stochastic} with $x$- and $t$-independent coefficients. We assume once again that there is no protein production ($k=0$) and that the  protein is stable ($\gamma=0$). In such a case, (\ref{unregulated_ME_Friedman_protein_partition_time_dependent_SI}) becomes the Master equation, describing 'jump process' between different states of the system (there is no drift term), and these 'jumps' are solely due to the cell division events. The probability that the system does not undergo such a 'jump', and that it is still in the same state at $t=\tau$ as it was at $t=0$, is equal to $\exp(-\Delta \tau)$ \cite{gardiner2009stochastic}. Therefore, the probability of a jump occurring in the infinitesimal interval $(t, t+dt)$ is given by \cite{van2007stochastic} 
\begin{equation}
\pi(t)dt = \Delta \exp(-\Delta t)dt.
\label{jump_probability_SI}
\end{equation}
%
%
%
In the case of a single lineage (scenario B mentioned above), we have $\Delta = \Delta_0$, and $\pi(t)$ must be identified with the cell cycle length distribution, $\pi(\tau)=f(\tau)$. Therefore, we obtain
\begin{equation}
f(\tau)=\Delta_0 \exp(-\Delta_0 \tau).
\label{generation_time_tau_final_form_SI_first_time}
\end{equation}
If we deal with the whole proliferating population, then $\Delta = 2 \Delta_0$ (scenario A), and $\pi(t)$ (\ref{jump_probability_SI}) reads
\begin{equation}
\pi(t) = 2\Delta_0 \exp(-2\Delta_0 t).
\label{jump_probability_SI_scenario_A}
\end{equation}
$\pi(\tau)$ (\ref{jump_probability_SI_scenario_A}) is identical with the quantity defined as 
\begin{equation}
\mathcal{C}(\tau) \equiv 2 e^{-\nu_m \tau} f(\tau) = 2\Delta_0 \exp(-2\Delta_0 \tau),
\label{Carrier_distribution_definition_SI}
\end{equation}
and called the 'carrier distribution' in \cite{powell1956growth}. 

\section{Cell age distribution $\phi(\tilde{a})$ \label{cell_age_distribution}}
To convince ourselves that (\ref{generation_time_tau_final_form_SI_first_time}) is valid, and to find the explicit form of the age distribution $\phi(\tilde{a})$, assume that $f(\tau) = \lambda \exp(-\lambda t)$, for $\lambda$ yet unspecified. 
From  Eq. (9) of Ref. \cite{powell1956growth} it follows that $\phi(\tilde{a})$ is given by 
%
\begin{equation}
\phi(\tilde{a}) = 2 \nu_m e^{-\nu_m}\int_{\tilde{a}}^{\infty} f(\tau) d\tau  = 2 \nu_m e^{-(\nu_m + \lambda)\tilde{a}}.
\label{age_distribution_first_time_SI}
\end{equation}
The normalization of $\phi(\tilde{a})$ (i.e., the condition $\int_{0}^{\infty}\phi(\tilde{a})d\tilde{a} =1$) yields $\nu_m = \lambda$, whereas from Eq. (\ref{growth_rate_equals_division_rate}) or from (\ref{evolution_mean_number_no_production_SI}) and (\ref{mean_protein_number_from_definitions_SI}) we obtain $\lambda = \Delta_0$, i.e., we recover Eq. (\ref{generation_time_tau_final_form_SI_first_time}). Hence, the age distribution $\phi(\tilde{a})$ is given by
\begin{equation}
\phi(\tilde{a}) = 2\Delta_0 e^{-2\Delta_0 \tilde{a}}.
\label{age_distribution_second_time_SI}
\end{equation}
%
%
%
It should be emphasized that functional forms of $f(\tau)$ (\ref{generation_time_tau_final_form_SI_first_time}), $\mathcal{C}(\tau)$ (\ref{Carrier_distribution_definition_SI}) and $\phi(\tilde{a})$ (\ref{age_distribution_second_time_SI}) are rather unrealistic. This is a consequence of a Poissonian nature of cell division in the present model. In particular, both $f(\tau)$ and $\mathcal{C}(\tau)$ should be unimodal, and vanishing not only for $\tau = 0$, but also for the values of $\tau$ sufficiently close to zero -- there certainly must be a minimal length of generation time. 

Because the functional forms of $f(\tau)$ (\ref{generation_time_tau_final_form_SI_first_time}) and $\mathcal{C}(\tau)$ (\ref{Carrier_distribution_definition_SI}) are identical, one can treat the results of numerical simulation of a single cell lineage as referring to the whole proliferating population, if only the division rate is rescaled. Namely, for a single call lineage (scenario B) we obtain identical protein number probability distribution as for the whole growing population (scenario A), provided that in the latter case the the true division rate $\Delta_0 = \nu_m \equiv \Delta_{0(A)}$ is two times smaller then in the former, i.e., $2\Delta_{0(A)} = \Delta_{0(B)}$.  

Note that the simulation curves shown in Fig. 2 in  the main text were generated by simulation of a single cell lineage with the cell division rate $\Delta_{0(B)}$ (scenario B, see Appendix \ref{subsec:simulation}). However, the theoretical curves $p_n(x; \omega(n), b)$ shown in Fig. 2 in the main text can be interpreted in two ways, depending on how we define the $\omega$ parameter: $\omega = k / \Delta_{0(A)}$ assumes Scenario A (whole population) and  $\omega = k / \Delta_{0(B)} = k / (2 \Delta_{0(A)})$ assumes Scenario B (single lineage). Therefore, the simulation results shown in Fig. 2 can also be reinterpreted as the results for Scenario A (whole population) where the cell division rate was twice smaller than the value set in the simulation algorithm: $\Delta_{0(A)} =  \Delta_{0(B)}/2$.


{The above discussion applies to} the batch culture. Analogous formulas can be derived for the continuous cell culture \cite{powell1956growth}, for which, in fact, conditions for state of balanced growth can be more easily reached and maintained.  

\section{Noise floor: Overview of other experimental results in the literature}\label{app:other_data}

A noise floor was observed in a number of experiments reported in the literature, however, these data were not suitable for comparison to our model. Below we overview the existing studies, which we are aware of, and which report the noise floor in gene expression.

In Ref. \cite{taniguchi2010quantifying}, the noise floor in gene expression in \textit{E. coli} manifested itself as a boomerang-shaped log-log plot of {the squared} coefficient of variation vs. mean gene expression. However, protein levels were measured in that Reference as concentrations and not absolute molecule numbers, which excluded the effect of protein loss at cell division inherent to our model. The description of the results in  Ref. \cite{taniguchi2010quantifying} may seem slightly misleading at first glance because the plots  in that reference were shown in the units of protein numbers. In fact, the protein fluorescence was measured in each cell and then its level for that cell was normalized by the volume of that same cell to get the protein concentration.  The protein concentrations were then again normalized by a mean cell  volume to obtain the description in the units of molecule numbers. And therefore, the resulting plots in Ref. \cite{taniguchi2010quantifying} show the protein numbers corresponding to the content of  average-sized cells, but the underlying method of measurement intentionally removed the effects of protein loss at cell division from the data. Thus, the results of Ref.  \cite{taniguchi2010quantifying}  should be interpreted using a model that describes protein levels in terms of concentrations. For that reason, the Friedman's model \cite{friedman2006linking} describing protein concentrations was used for data fitting in Ref.  \cite{taniguchi2010quantifying}. Our model cannot be fitted to these data because it describes protein levels in terms of protein numbers and thus it explicitly accounts for protein loss at cell division.

In Refs. \cite{Silander2012} and \cite{Wolf2015}, a noise floor was observed in gene expression in \textit{E. coli}. These data were unsuitable for comparison to our model because they were gated to observe the gene expression levels in only those cells that were in similar cell-cycle stages. Thus, the effect of protein loss due to cell division would not be visible in these data. A noise floor was also observed in gated data in \textit{S. cerevisiae} \cite{newman2006single, bar2006noise}.

\section{Distribution tails \label{Tails}}
%
%
%
The protein number distribution $p(x)$ as given by Eqs. (\ref{p_d_as_infinite_sum}) and (\ref{D_i_p_d_as_infinite_sum_eta_d_iloczyn_stochastic_production}) in the main text, i.e.,
\begin{equation}
p(x)=\sum^{\infty}_{i=0} \frac{C_i(\omega)}{b} \exp\left(\frac{-2^{i}x}{b(\omega+1)} \right),
\label{p_d_as_a_infinite_sum_SI}
\end{equation}
where 
\begin{equation}
C_{i}(\omega) = \frac{2^{i} \omega}{(1+\omega)^2}  \frac{(2(1+\omega)^{-1};2)_i}{(2;2)_i}\frac{(\frac{1}{2}(1+\omega)^{-1};\frac{1}{2})_{\infty}}{(\frac{1}{2};\frac{1}{2})_{\infty}},
\label{D_i_of_p_d_as_a_infinite_sum_for_eta_d_iloczyn_for_stochastic_production_SI}
\end{equation}
has exponential tail of the form $\exp\{-x/[b(\omega+1)]\}$ (the leading exponent in (\ref{p_d_as_a_infinite_sum_SI})). The same is true for $p_u(x)$ (\ref{p(x)_for_eta_u_SI_2}) (see the next section), i.e. the solution of Eq. (\ref{unregulated_ME_Friedman_protein_partition_ssSI}) for $\eta(q) = \Theta (q)\Theta \left( 1-q \right) = 1$ given by Eq. (\ref{eta_U}). In fact, it can be shown that for most of functional forms of $\eta(q)$, the solution of Eq. (\ref{unregulated_ME_Friedman_protein_partition_ssSI}) has exactly the same exponential tail (there are, however, exceptions: for example, $\eta_{}(q)=\frac{1}{2}\left[\delta (q) + \delta (1-q) \right]$ yields the tail of the form $\exp\{-x/[b(2\omega+1)]\}$). 

This is a special case of a yet more general result. We assume here that $g(x)=0$ and the burst size PDF is exponential, as given by Eq. (\ref{exponential_nu}), 
\begin{equation}
\nu_{}(u) = \frac{e^{-u/b}}{b}, ~~~~  \mathcal{L}[\nu_{}(u)] \equiv \hat{\nu}(s) = 1/(sb+1),
\label{exponential_nu_SI}
\end{equation}
but we allow for arbitrary $x$-dependence of $h_d(x)$ and $h_p(x)$ in Eq. (\ref{unregulated_t_dependent_ME_of_Friedman_with_protein_partition_PBE}). Following Ref. \cite{friedlander2008cellular}, we neglect the $q$-integral term in the steady-state limit of Eq. (\ref{unregulated_t_dependent_ME_of_Friedman_with_protein_partition_PBE}) (for large protein number $x$, the contribution of states with still larger $x$ may be neglected) and obtain   
\begin{eqnarray}
0 &=& - \left[ h_d(x) + \int_0^{\infty} h_d(\zeta) p(\zeta) d \zeta \right] p(x)\nonumber \\
&+& k \int_{0}^{x} w(x-x^{\prime})p(x^{\prime}) dx^{\prime}.
\label{unregulated_t_dependent_ME_of_Friedman_with_protein_partition_tails}
\end{eqnarray}  
Next, we combine the derivative of (\ref{unregulated_t_dependent_ME_of_Friedman_with_protein_partition_tails}) with respect to $x$ with the original equation and obtain
\begin{eqnarray}
\frac{p^{\prime}(x)}{p(x)} = \frac{k h^{\prime}_p(x) + h^{\prime}_d(x) + h_d(x)/b + R/b}{k h_p(x) + h_d(x) + R}, \\ 
\label{general_ODE_for_tails}
\end{eqnarray}
where 
%
\begin{equation}
R=\int_0^{\infty} h_d(\zeta) p(\zeta) d \zeta = \nu_m,
\label{definition_of_R}
\end{equation}
c.f. Eq. (\ref{population_growth_rate_Mantzaris}). From (\ref{general_ODE_for_tails}) it follows that at the tail of the PDF  the ratio  $p(x_2)/p(x_1)$ is well approximated by 
\begin{eqnarray}
\mathcal{P}_{12} = \exp \left(- \mathcal{I}_{12} \right),
\label{solution_of_general_ODE_for_tails_SI_Exponent}
\end{eqnarray}
where
\begin{equation}
\mathcal{I}_{12} = \int_{x_1}^{x_2}
\frac{k h^{\prime}_p(x) + h^{\prime}_d(x) + h_d(x)/b + R/b}{k h_p(x) + h_d(x) + R} dx.
\label{solutionofgeneralODEfortailsSIIntegral}
\end{equation}
For $h_d(x) = \Delta/2$ and $h_p(x)=1$, from (\ref{definition_of_R}), (\ref{solution_of_general_ODE_for_tails_SI_Exponent}) and (\ref{solutionofgeneralODEfortailsSIIntegral}) we obtain the tail of the form $\exp\{-x/[b(\omega+1)]\}$.

Note that the the above derivation is not universally valid. For example, for the 'all or none' mode of protein partitioning, i.e., for $\eta_{b}(q)=\left[\delta (q) + \delta (1-q) \right]/2$, one cannot simply drop out the integral over $q$ in Eq. (\ref{unregulated_t_dependent_ME_of_Friedman_with_protein_partition_PBE}). However, apart from such rather pathological situations, it seems that the derivation of the large-$x$ asymptotic behaviour of $p(x)$ proposed in Ref. \cite{friedlander2008cellular} can be generalized to the present case of stochastic protein production, provided the burst size PDF is of the form (\ref{exponential_nu_SI}) and there are is no protein decay.

\section{Distribution of protein numbers for the uniform protein partition ratio distribution. \label{protein_PDF_for_eta_u}}
%

In this section we solve Eq. (\ref{unregulated_ME_Friedman_protein_partition_ssSI}) (Eq. (2) of the main text) for $\gamma=0$, exponential burst size PDF as given by Eq. (\ref{exponential_nu_SI}), and for uniform partition ratio distribution  
%
\begin{eqnarray}
\eta(q) = \Theta (q)\Theta \left( 1-q \right) = 1,
\label{eta_U}
\end{eqnarray}
for $0<q<1$, where $\Theta (q)$ denotes the  Heaviside theta function. $\eta(q)$ (\ref{eta_U}) is not very realistic at large $x$, where any such distribution should be peaked around $q=1/2$, like, e.g., the symmetric beta distribution (a continuous counterpart of the binomial distribution), considered in Ref. \cite{mantzaris2006stochastic}. $\eta(q)$ (\ref{eta_U}) becomes more acceptable for $x$ of order of few protein molecules. Yet, $\eta(q)$ (\ref{eta_U}) has been used in Refs. \cite{friedlander2008cellular, brenner2007nonequilibrium}, probably due to its mathematical simplicity -- it is one of the few examples of a partition ratio PDF, for which analytical solutions of PBE-like equations are known. 

For $\gamma=0$ and $\eta(q)$ (\ref{eta_U}), Eq. (3) reads  
%
\begin{eqnarray}
0 &=& k\hat{w}(s) G(s) + \Delta \int_{0}^{1} [G(qs) - G(s)]dq.
\label{unregulated_ME_of_Friedman_in_s_space_ss_for_eta_u}
\end{eqnarray}
By differentiating (\ref{unregulated_ME_of_Friedman_in_s_space_ss_for_eta_u}) with respect to $s$ and combining such obtained equation with the original one, integrating by parts and using the identity 
\begin{eqnarray}
\int_{0}^{1} q G^{\prime}(qs) dq = \frac{1}{s} \left[G(s) - \int_{0}^{1} G(qs)dq \right], 
\label{identity_for_getting_rid_of_derivative_for_uniform_eta}
\end{eqnarray}
in order to to cancel the terms containing $G(qs)$ (note that $G^{\prime}(qs) \equiv (d G(y)/dy)_{y =q s}$), we finally get
\begin{eqnarray}
G^{\prime}(s) &=& \frac{\omega \left[s\hat{w}^{\prime}(s) + \hat{w}(s) \right]}{s[1-\omega\hat{w}(s)]} G(s).
\label{unregulated_ME_of_Friedman_in_s_space_ss_for_eta_u_ode}
\end{eqnarray}
ODE (\ref{unregulated_ME_of_Friedman_in_s_space_ss_for_eta_u_ode}) is equivalent to the integral equation (\ref{unregulated_ME_of_Friedman_in_s_space_ss_for_eta_u}). For $\hat{\nu}(s)$ (\ref{exponential_nu_SI}) we have $\hat{w}(s)=\hat{\nu}(s)-1= -sb/(sb+1)$ and Eq. (\ref{unregulated_ME_of_Friedman_in_s_space_ss_for_eta_u_ode}) has the following solution:
\begin{equation}
G(s) = (1-\epsilon)\frac{\xi^{\epsilon}}{(s+\xi)^{\epsilon}} + \epsilon\frac{\xi^{\epsilon+1}}{(s+\xi)^{\epsilon+1}},
\label{G(s)_for_eta_u_SI}
\end{equation}
where $\epsilon = \omega/(\omega+1)$, $\xi = 1/[b(\omega+1)]$.
%
The inverse Laplace transform of $G(s)$ (\ref{G(s)_for_eta_u_SI}) reads 
\begin{equation}
p_u(x) =(1-\epsilon)\frac{\xi^{\epsilon} x^{\epsilon-1} e^{-\xi x}}{\Gamma(\epsilon)} + \epsilon\frac{\xi^{\epsilon+1} x^{\epsilon} e^{-\xi x}}{\Gamma(\epsilon+1)}.
\label{p(x)_for_eta_u_SI_2}
\end{equation}
Because $p_u(x)$ (\ref{p(x)_for_eta_u_SI_2}) is a statistical mixture of two broad  gamma distributions (note that $0 < \epsilon < 1$), $p_u(x)$ itself is broad ($c^2_v \geq 1/2$). 

For the case of deterministic protein production, instead of $p_u(x)$ (\ref{p(x)_for_eta_u_SI_2}) we obtain a gamma distribution \cite{friedlander2008cellular, brenner2007nonequilibrium}, cf. next Section.

\section{Distribution of protein  numbers for the deterministic protein production and half-by-half protein partitioning \label{det_prot_prod}}
We consider here the case of unregulated, deterministic protein production, i.e., in Eq. (\ref{unregulated_t_dependent_ME_of_Friedman_with_protein_partition_PBE}) we put $k=0$ and $g(x) = \sigma > 0$. For $h_d(x) = \Delta/2$, in the steady-state limit instead of Eq. (\ref{unregulated_ME_Friedman_protein_partition_ssSI}) (Eq. (2)) we obtain 
\begin{equation}
\lambda  \frac{d}{d x}\left[p(x)\right] = \int_{0}^{1} \eta(q)\left[\frac{1}{q} p\left(\frac{x}{q}\right) - p(x)\right] dq,
\label{unregulated_ME_of_Friedman_with_protein_partition_ss_det_prod_SI}
\end{equation}
where $\lambda = \sigma/\Delta$. Laplace transform of Eq. (\ref{unregulated_ME_of_Friedman_with_protein_partition_ss_det_prod_SI}) yields
\begin{equation}
\lambda  sG(s) =  \int_{0}^{1} \eta(q) \left[G(qs)- G(s)\right] dq,
\label{unregulated_ME_of_Friedman_in_s_space_ss_det_prod_SI}
\end{equation}
and we have assumed $p(0{+}) \equiv \lim_{x \to 0{+}} p(x) = 0$ (validity of this assumption is checked \textit{a posteriori}). For the half-by-half protein partitioning, $\eta(q)=\delta(q-\frac{1}{2})$, Eq. (\ref{unregulated_ME_of_Friedman_in_s_space_ss_det_prod_SI}) can be rewritten as 
\begin{equation}
G(s) = \left(1 + \lambda s\right)^{-1} G\left(\frac{s}{2} \right).
\label{R(s)_for_eta_d_iloczyn_for_deterministic_production}
\end{equation}
%
%
%
Eq. (\ref{R(s)_for_eta_d_iloczyn_for_deterministic_production}) can be solved by iteration; we obtain
\begin{equation}
G(s) = \prod_{i=0}^{\infty} \frac{1}{1+2^{-i}\lambda s} = \frac{1}{(-\lambda s;\frac{1}{2})_{\infty}} = \sum_{r=0}^{\infty} \frac{(-\lambda s)^r}{(\frac{1}{2};\frac{1}{2})_{r}},
\label{unregulated_ME_of_Friedman_in_s_space_ss_for_eta_d_iloczyn_for_deterministic_production}
\end{equation}
where $(a;q)_k$ is a $q$-Pochhammer Symbol. In (\ref{unregulated_ME_of_Friedman_in_s_space_ss_for_eta_d_iloczyn_for_deterministic_production}) we have used the following identity
\begin{equation}
\frac{1}{(z;q)_{\infty}} =
\sum_{n=0}^{\infty}\frac{z^n}{(q;q)_n}, 
\label{q_Binomial_Theorem_SI}
\end{equation}
which is a special case of the $q$-binomial theorem \cite{andrews1986qseries}. Note that $G(s)$ (\ref{unregulated_ME_of_Friedman_in_s_space_ss_for_eta_d_iloczyn_for_deterministic_production}) can be also rewritten as
\begin{eqnarray}
G(s) &=& \sum_{r=0}^{\infty} \frac{(-2\lambda s)^r}{\left[r\right]_{\frac{1}{2}}!} = e_{\frac{1}{2}}(-2\lambda s),
\label{unregulated_ME_of_Friedman_in_s_space_ss_for_eta_d_iloczyn_for_deterministic_production_bis_more_math}
\end{eqnarray}
where $\left[r\right]_{q}! \equiv (1-q)^{-r} (q;q)_r $ denotes the $q$-factorial and $e_{q}(x)$ is the $q$-exponential function.

The inverse Laplace transform of (\ref{unregulated_ME_of_Friedman_in_s_space_ss_for_eta_d_iloczyn_for_deterministic_production}) reads
\begin{equation}
p_{}(x; \lambda)= \frac{1}{\lambda} \sum^{\infty}_{i=0}  D_{i} \exp\left(-\frac{2^{i}x}{\lambda} \right),
\label{p_d_as_a_infinite_sum_for_eta_d_iloczyn_for_deterministic_production}
\end{equation} 
where the coefficients $D_{i}$ are given by
\begin{equation}
D_{i} =  \prod^{i}_{j=1} \frac{2}{1-2^j} \prod^{\infty}_{l=1} \frac{2^l}{2^l-1} = \frac{2^{i}}{(2;2)_i}\frac{1}{(\frac{1}{2};\frac{1}{2})_{\infty}}.
\label{D_i_of_p_d_as_a_infinite_sum_for_eta_d_iloczyn_for_deterministic_production}
\end{equation}
\begin{figure}
\begin{center}					  				
\rotatebox{270}{\scalebox{0.3}{\includegraphics{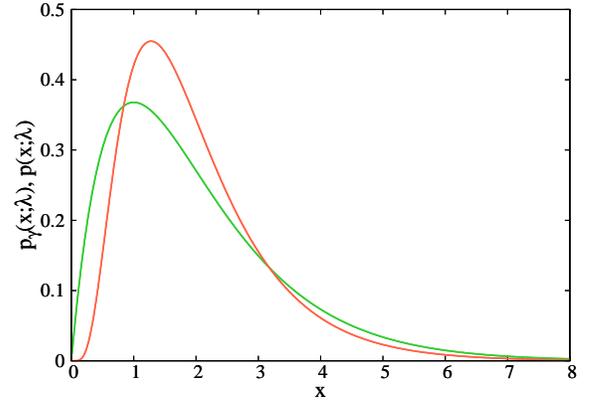}}} 
\end{center}  
\caption{Protein  number PDFs of our model in case of deterministic protein production. $p_{}(x; \lambda)$ (\ref{p_d_as_a_infinite_sum_for_eta_d_iloczyn_for_deterministic_production}) (red) is a solution of Eq. (\ref{unregulated_ME_of_Friedman_in_s_space_ss_det_prod_SI}) for $\eta(q)=\delta(q-1/2)$ (\ref{half_by_half_eta}), whereas the gamma distribution $p_{\gamma}(x; \lambda)=\lambda^{-2} x \exp(-x/\lambda)$ (\ref{gamma_for_det_prod_and_eta_U}) (green) is a solution of Eq. (\ref{unregulated_ME_of_Friedman_in_s_space_ss_det_prod_SI}) for $\eta(q)=\Theta (q)\Theta \left( 1-q \right) $ (\ref{eta_U}); both are plotted here for $\lambda=1$, i.e., as a function of  the rescaled variable $x/\lambda$. Both distributions have the same average value, $\langle x \rangle = \mu_1 = 2\lambda$, and identical exponential tails $\sim \exp(-x/\lambda)$ but seemingly different shapes. The variance of $p_{}(x; \lambda)$, equal to $\frac{4}{3}\lambda^2$, is smaller than the variance of $p_{\gamma}(x; \lambda)$, equal to $2\lambda^2$. In contrast to $p_{\gamma}(x; \lambda)$, the maximum of $p_{}(x; \lambda)$ cannot be found analytically; for $\lambda=1$ we have $x_m \approx 1.2773$ and $p(x_m) \approx 0.4549$. Note that, for both $p_{}(x; \lambda)$  and $p_{\gamma}(x; \lambda)$, the {squared} coefficient of variation is constant (does not depend on $\lambda$, i.e., on the mean protein number), and is equal to its minimal value for the corresponding solution with stochastic protein production.}
\label{detprod_fig_eta_uni_and_eta_det} 
\end{figure}
For $i=0$, we obtain the leading-order term, i.e., the slowest-decaying exponent, $e^{- x/\lambda}$, which determines the behaviour of $p_{}(x; \lambda)$ (\ref{p_d_as_a_infinite_sum_for_eta_d_iloczyn_for_deterministic_production}) at the $x\to \infty$ limit. Note that $p_{}(x; \lambda)$ (\ref{p_d_as_a_infinite_sum_for_eta_d_iloczyn_for_deterministic_production}) can be regarded as a counterpart of the probability distribution given by (\ref{p_d_as_a_infinite_sum_SI}) and (\ref{D_i_of_p_d_as_a_infinite_sum_for_eta_d_iloczyn_for_stochastic_production_SI}) (Eqs. (14) and (15) in the main text) in the case of deterministic protein production. 


From $G(s)$ (\ref{unregulated_ME_of_Friedman_in_s_space_ss_for_eta_d_iloczyn_for_deterministic_production}) we obtain both the moments ($\mu_r$) and the cumulants ($\kappa_r$) of $\mathcal{L}^{-1}[G(s)] \equiv p(x; \lambda)$ (\ref{p_d_as_a_infinite_sum_for_eta_d_iloczyn_for_deterministic_production}), namely 
\begin{equation}
\mu_r = \frac{\lambda^r r!}{(\frac{1}{2};\frac{1}{2})_{r}} = \frac{2^r \lambda^r r! }{\left[r\right]_{\frac{1}{2}}!},  
\label{moments_for_eta_d_and_deterministic_production}
\end{equation}
\begin{equation}
\kappa_r =  \frac{2^r \lambda^r (r-1)!}{2^r-1}.  
\label{cumulants_for_eta_d_and_deterministic_production}
\end{equation}
In the present case, $\mu_r $ can be also easily obtained from moments equations.

For a uniform distribution of protein division ratio (\ref{eta_U}), the solution of Eq. (\ref{unregulated_ME_of_Friedman_with_protein_partition_ss_det_prod_SI}) reads \cite{friedlander2008cellular} 
\begin{eqnarray}
p_{\gamma}(x;\lambda) &=& \lambda^{-2} x e^{-x/\lambda}.
\label{gamma_for_det_prod_and_eta_U}
\end{eqnarray}
The gamma distribution $p_{\gamma}(x;\lambda)$ (\ref{gamma_for_det_prod_and_eta_U}) is therefore a deterministic counterpart of $p_{u}(x)$ (\ref{p(x)_for_eta_u_SI_2}).

Both $p_{}(x; \lambda)$ (\ref{p_d_as_a_infinite_sum_for_eta_d_iloczyn_for_deterministic_production}) and $p_{\gamma}(x; \lambda)$ (\ref{gamma_for_det_prod_and_eta_U}) define a one-parameter family of probability distributions; we have $p_{}(x, \lambda) = \lambda^{-1} f_{}(x/\lambda)$, where $f(z) = p_{}(z, 1)$ is a function of the rescaled variable $x/\lambda$. Also, both $p_{}(x; \lambda)$ and $p_{\gamma}(x; \lambda)$ are right-skewed, unimodal, and vanish in the $x \to 0{+}$ limit, as has been assumed for $p_{}(x; \lambda)$, cf. Fig \ref{detprod_fig_eta_uni_and_eta_det}. However, in contrast to  the gamma distribution, for $p_{}(x; \lambda)$  we have $p^{(k)}_{}(0; \lambda)=0$ for arbitrary $k$, hence $p_{}(x; \lambda)$ is not analytic at $x=0$. In consequence, the Taylor expansion of $p_{}(x; \lambda)$  at $x_0=0$ does not exists. This can be shown  using properties of $G_{}(s)$ (\ref{unregulated_ME_of_Friedman_in_s_space_ss_for_eta_d_iloczyn_for_deterministic_production}). We have 
\begin{eqnarray}
p(0) &=& \lim_{s \to \infty} s G_{}(s) = 0, \nonumber \\
p^{\prime}(0) &=& \lim_{s \to \infty} s \left[ s G_{}(s) - p(0)\right] = 0, \nonumber \\
p^{\prime \prime}(0) &=& \lim_{s \to \infty} s \left[ s^2 G_{}(s) - s p(0) - p^{\prime}(0) \right] = 0, \nonumber \\
& & \hdots
\end{eqnarray}
etc., due to the obvious relation 
\begin{eqnarray}
& & \lim_{s \to \infty} s^k G_{}(s) = 0,
\end{eqnarray}
valid for any $k<\infty$.




\section{Similarity of gamma distributions and the distributions resulting from our model} 

\begin{figure}
\rotatebox{0}{\scalebox{0.3}{\includegraphics{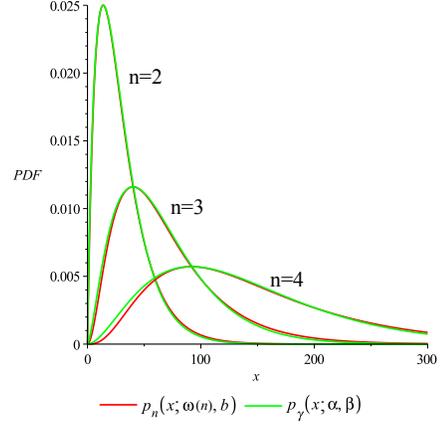}}}  
\caption{Gamma distributions $p_\gamma(x;\alpha, \beta)$ (\ref{gamma_distribution_definition_alfa_beta_SI}) are almost indistinguishable from the $p_n(x;\omega{(n)}, b)$ distributions of our model as given by Eq. (\ref{p_d_as_a_finite_sum_specyfic_value_of_omega_SI}), when fitted by matching their maxima. Red lines: $p_n(x;\omega{(n)}, b)$ with $b=5$ and $n=2,3,4$. Green lines: The corresponding $p_\gamma(x;\alpha, \beta)$ (\ref{gamma_distribution_definition_alfa_beta_SI}), whose parameter values are given in Table \ref{table_fitSI_labels}.}
\label{fitSI_labels} 
\end{figure}

In this section we present an alternative method of fitting of gamma distributions to the distributions resulting from our model: We match the maxima of our distribution, 
\begin{eqnarray}
p_n(x;\omega{(n)}, b) &=& \frac{1}{b} \sum^{n}_{l=1} A_{n,l} \exp\left(-\frac{x}{2^l b } \right),
\label{p_d_as_a_finite_sum_specyfic_value_of_omega_SI}
\end{eqnarray}
where $A_{n,l} = (-1)^{n-l} 2^{\frac{l(l-3)}{2}}/[\prod^{l-1}_{i=1} (2^i-1)\prod^{n-l}_{j=1} (2^j-1)]$ 
(c.f. Eqs. (\ref{p_d_finite_sum_specyfic_value_omega}) and (\ref{Definition_of_A_n_l}) of the main text) and of the gamma distribution $p_\gamma(x;\alpha, \beta)$, c.f. Eq. (\ref{gamma_distribution_definition_alfabeta}) of the main text, i.e.:
\begin{equation}
p_{\gamma}(x; \alpha, \beta) \equiv \frac{x^{\alpha-1} \exp\left(-x/\beta \right)}{\beta^{\alpha}\Gamma(\alpha) }.
\label{gamma_distribution_definition_alfa_beta_SI}
\end{equation}
The coordinates of the gamma distribution's maximum for $\alpha>1$ are given by:
\begin{eqnarray}
 x_{max,p_{\gamma}} &=& \beta\,\alpha-\beta, \nonumber \\
y_{max,p_{\gamma}} &=& {\frac {   \left[  \left( \alpha-1
 \right) \beta \right] ^{\alpha-1}{\exp(1-\alpha)}}{{\beta}^{\alpha}   \Gamma 
 \left( \alpha \right) }}.
\end{eqnarray}
The coordinates $x_{max,p_{n}}$ and $y_{max,p_{n}}$ of the maximum of $p_n(x;\omega{(n)}, b)$ can be found analytically for small values of $n$ (in our case, the Maple software was able to find them explicitly for $n \leq 5$). Otherwise, they can be calculated numerically. In Fig. \ref{fitSI_labels}, we put $b=5$ and $n=2,3,4$ (corresponding to $\omega=3,7,15$) in  $p_n(x;\omega{(n)}, b)$. By numerically solving the equations 
\begin{eqnarray}
 x_{max,p_{n}}&=&x_{max,p_{\gamma}}, \nonumber\\
 y_{max,p_{n}}&=&y_{max,p_{\gamma}},
\end{eqnarray}
we found the $\alpha$ and $\beta$ parameters of the gamma distributions whose maxima exactly match the maxima of $p_n(x;\omega{(n)}, b)$. 

Fig. \ref{fitSI_labels} shows that this way of fitting yields the gamma distribution plots that are even more similar to $p_n(x;\omega{(n)}, b)$ than those obtained by matching the first two moments of the distributions, as in the main text, Fig 2. 
\setlength{\tabcolsep}{5pt}
\begin{table}[h]
\begin{center}
\begin{tabular}{l| l l l}
$n$       & 2    & 3 & 4\\
\hline
$\omega$  & 3    & 7 & 15\\
\hline
$\alpha$  & 1.90 & 2.52 & 2.88\\
\hline
$\beta/b$ & 3.07 & 5.28 & 9.74
\end{tabular}
\end{center}
\caption{Parameters of the distributions $p_n(x;\omega{(n)}, b)$ and $p_\gamma(x;\alpha, \beta)$ shown in Fig. \ref{fitSI_labels}.}\label{table_fitSI_labels}
\end{table}

\section{A property of the distribution $p_n(x)$ }\label{subsec:property}
We denote the Laplace transform of $p_n(x)$ (\ref{p_d_as_a_finite_sum_specyfic_value_of_omega_SI}) as $G_n(s)$. From Eq. (12) of the main text, it follows that
\begin{eqnarray}
G_n(s)&=&  \frac{2^{-\frac{n(n+1)}{2}} b^n}{\left(s + \frac{b}{2}\right)\left(s + \frac{b}{4}\right)\ldots \left(s + \frac{b}{2^n}\right)}.
\end{eqnarray}
One can show that  
\begin{eqnarray}\label{eq:G}
s \ G_{n+1}(s) = \frac{b}{2^{n+1}}\left[ G_n(s) - G_{n+1}(s) \right].
\end{eqnarray}
Assuming
\begin{eqnarray}\label{eq:p0}
p_{n+1}(0)=0 \ \ \mathrm{for} \ n>0,  
\end{eqnarray}
it follows from Eq. (\ref{eq:G}) that
\begin{eqnarray}\label{eq:p}
\frac{d}{dx}  p_{n+1}(x) = \frac{b}{2^{n+1}} \left[ p_n(x) - p_{n+1}(x) \right].
\end{eqnarray}
We can solve Eq. (\ref{eq:p}) iteratively for $n\geq 1$ to obtain the formulas  for consecutive $p_{n+1}(x)$, starting from $p_1(x) = (b/2)\exp(-b x /2)$ and using (\ref{eq:p0}) as the boundary condition. From Eq. (\ref{eq:p}) it follows that $p_{n+1}(x)$ has a maximum in the point where its plot intersects with the plot of $p_{n}(x)$. 

\section{Simulation}\label{subsec:simulation}

The simulation results shown in Fig. 2 in the main text were obtained using a custom extension of the StochPy package \cite{maarleveld2013stochpy}: The standard simulation using the direct Gillespie algorithm was supplemented with cell division. Histograms were calculated along a single trajectory, which mimics the observation of a single cell lineage. The initial part of the trajectory was cut off to obtain only the steady-state behavior. The reaction kinetics (see the file \texttt{NonRegulatedGeneCellDivision.psc}) was given by:
\begin{eqnarray}
&\ce{O  ->[k] O + M}& \\ \nonumber
&\ce{ M ->[k_p] M + P}& \\ \nonumber
&\ce{M  ->[k_{dm}] \varnothing}&
\end{eqnarray}
where \ce{O} denotes the gene promoter, \ce{M} is mRNA, and \ce{P} is protein. $k$ is transcription rate and $k_p$ is translation rate. $k_{dm}$ denotes mRNA degradation rate and its value is chosen such that mRNA life time is much shorter than the mean cell cycle duration $1/\Delta_{0{(B)}}$, consistently with the assumption of the analytical model described in the main text. Additionally, cell division occurs at the rate $\Delta_{0{(B)}}$: The custom function \texttt{multiple\_division()} draws a random cell cycle length $T$ from a specified distribution (here: exponential with the mean $1/\Delta_{0{(B)}}$). In a given cell cycle, the simulation runs until the cell age $T$ and at that time point  the molecule numbers $M$ and $P$ are divided by 2. If the remainder of the division is 1, then the remaining molecule goes to the "observed" daughter cell with probability 1/2. The simulation for the next cell cycle is initialized with the resulting molecule numbers $M$ and $P$ for the daughter cell. Simulation parameter values are shown in Table \ref{table_params} and  \ref{table_params2}. The simulation code, containing the custom functions that implement cell division can be found in the file \texttt{NonRegulatedGeneCellDivision.ipynb} (Jupyter notebook file, Supplemental Material) \footnote{See Supplemental Material at [URL will be inserted by publisher] for the simulation code.}.

\begin{table}[h]
	\begin{center}
	\begin{tabular}{l l}	
		$\Delta_{0{(B)}}$ &  1\\ 
		\hline 
		$k_p$ & 250 \\ 
		\hline 
	    $k_{dm}$& 50 \\
	    \hline 
	    {\scriptsize \texttt{max\_simulation\_time}} & 20000 \\
	 	\hline 
	    {\scriptsize \texttt{time\_cutoff}} & 2000
    \end{tabular} 
	\end{center}
	\caption{Simulation parameter values for Fig. 2 in the main text. These values were the same in all simulations.}\label{table_params}
\end{table}

\begin{table}[h!]
	\begin{center}
		\begin{tabular}{l | l l l l}
			$n$ & 1 & 2 & 3 & 4 \\
			\hline
			$k$ & 1 & 3 & 7 & 15 
		\end{tabular} 	
	\end{center}
	\caption{Simulation parameter values for Fig. 2 in the main text. These values correspond to the specific curves in Fig. 2.}\label{table_params2}
\end{table}
%


\bibliography{bibliography25a}


\end{document}